\documentclass[fleqn,10pt]{wlscirep}
\usepackage[utf8]{inputenc}
\usepackage[T1]{fontenc}
\title{Developments of a centimeter-level precise muometric wireless navigation system (MuWNS-V) and its first demonstration using directional information from tracking detectors}

\author[1,2,*]{Dezso Varga}
\author[1,3]{Hiroyuki K. M. Tanaka}
\affil[1]{The International Virtual Muography Institute, global}
\affil[2]{Wigner Research Centre for Physics, Budapest, Hungary}
\affil[3]{The University of Tokyo, Tokyo, Japan}

\affil[*]{Varga.Dezso@wigner.hu}

\begin{abstract}
Various positioning techniques such as Wi-Fi positioning system have been proposed to use in situations where satellite navigation is unavailable. One such system, the muometric positioning system (muPS), was invented for navigation which operates in locations where even radio waves cannot reach such as underwater or underground. muPS takes advantage of a key feature of its probe, cosmic-ray muons, which travel straightforwardly at almost a speed of light in vacuum regardless of the matter they traverse. Similar to other positioning techniques, muPS is a technique to determine the position of a client's muPS receiver within the coordinate defined by reference detectors. This can be achieved either by using time-of-flight (ToF) or angle of arrival (AoA) measurements. The latter configuration (AoA), called the Vector-muPS has recently been invented and the present paper describes the developments of the first prototype of a vector muometric wireless navigation system (MuWNS-V) with this new vector-muPS concept and its demonstration. With MuWNS-V, the reference tracker and the receiver ran wirelessly with fully independent readout systems, and a positioning accuracy of 3.9 cm (RMS) has been achieved. We also evaluated the outcome of measuring continuous indoor localization of a moving receiver with this prototype. Our results indicated that further improvements in positioning accuracy will be attainable by acquiring higher angular resolution of the reference trackers. It is anticipated that "sub-cm level" navigation will be possible for muPS which could be applied to many situations such as future autonomous mobile robot operations. 
\end{abstract}
\begin{document}

\flushbottom
\maketitle
\thispagestyle{empty}

%Please note: Abbreviations should be introduced at the first mention in the main text – no abbreviations lists. Suggested structure of main text (not enforced) is provided below.

\section*{Introduction}

Thus far various indoor positioning systems have been invented and demonstrated. While Wi-Fi indoor positioning system (IPS) \cite{Mapsted_2023} is commonly used, one of the most successful state-of-the-art radio-wave techniques are those using radio waves is Marvelmind Indoor “GPS” \cite{Marvelmind_2023} that enables a positioning accuracy of 2 cm which offers a similar level or better performance than GPS-RTK \cite{Trimble_2023}. However, if there are radio frequency shielding obstacles such as metal or seawater between the access points and the client receiver devices, their navigation probes are blocked and as a result, the navigation quality will degrade significantly or the navigation signal itself will be unavailable. LiDAR \cite{Chow_2019}, Dead reckoning \cite{Khedr_2021}, photographic-based imagery \cite{Pan_1995} can also be used for specific purposes of navigation such as the ones used in the collision avoidance support system \cite{Lidar_2023}, however, since these techniques do not provide coordinate values of the clients within the given coordinate, navigation routes cannot be programmed with these methods. On the other hand, navigation routes can be programmed by utilizing dead reckoning \cite{Khedr_2021}. This is a method of estimating past and present positions based on the route traveled, distance traveled, starting point, drift, etc., and performing navigation based on that position information. An inertial measurement unit (IMU) is a type of a device used for dead reckoning navigation with inertial sensors (an integrated package of accelerometers and gyroscopes), the positioning update rate is usually high and the navigation accuracy is not affected by obstacles or other interference in the surrounding environments. However, even with an industrial level IMU, the positioning estimation error increases to more than 50 m within one minute mainly due to the gyroscope’s bias instability and angle random walk \cite{Borodacz_2022}. muPS was invented to address the aforementioned problems and to operate in situations where other positioning methods may fail or malfunction.

Cosmic-ray muons are elementary particles, created by natural phenomena in the upper atmosphere. Muons with energies much higher than natural radioactivity reach the Earth surface and subsurface which can be utilized for a number of possible applications including imagery \cite{Tanaka_2014},\cite{Morishima_2017},\cite{Varga_2020}, navigation \cite{Tanaka_mups},\cite{Tanaka_wns},\cite{Tanaka_wns2}, time metrology \cite{Tanaka_cts},\cite{Tanaka_cts2},\cite{Tanaka_ctc}, and cryptographic communication \cite{Tanaka_coding},\cite{Tanaka_coding2}. The main properties of cosmic muons have been known for decades, including precise information on their abundance (flux), energy distribution, and properties when interacting with material \cite{Olah_agu}.

A method to use cosmic-ray muons to determine the three-dimensional position of a receiver (muon detector), the "muPS" was introduced in 2020 by Tanaka \cite{Tanaka_mups}. For many applications, it is highly advantageous to have the option to use wireless, asynchronous communication between the reference system (in a known location), and the receiver (in an unknown location) for positioning. The apparatus system that enables muometric navigation is called the Muometric Wireless Navigation System (MuWNS) \cite{Tanaka_wns},\cite{Tanaka_wns2}. However, the previous iterations of muPS required either (A) wires between the reference and the client’s receivers or (B) precise and stable clocks associated with the reference and the receivers. The restriction of (A) substantially degrades the flexibility of muPS operation, and that of (B) requires atomic clocks with extraordinarily high granularity, such as an optical pumping cesium clock, for stable and acceptable navigation accuracy levels (< 10 cm) \cite{Tanaka_wns2}.  For example, if it takes 10 seconds in order to collect four muon tracks for determining x, y, z, and t, granularity of these clocks must be better than 300 ps in 10 s. On the other hand, short-term stabilities (RMS) of a commercially available Rb chip scale atomic clock (CSAC) is 1 ns in 10 s \cite{Microchip_2023}. These problems can be solved with Vector muPS, a concept which was proposed by Tanaka in 2023 \cite{Tanaka_vmups}. The Vector-muPS concept consists of the following key elements:

\begin{itemize}

\item{A Reference (with a known position and orientation) is equipped with a good angular resolution, high efficiency, and possibly large-scale (large aperture) muon tracking system.}

\item{A Receiver (with unknown position) is the device that will be positioned by the Reference. The principal objective is to determine the location of the Receiver with sufficient precision, within a practically convenient time, and to follow its motion if it is not stationary.}

\item{One can assume that there is some limited, wireless, non-synchronized communication channel between the Receiver and the Reference, e.g. WiFi, ultrasonic, infrared or light-based data transfer.}

\end{itemize}

The present paper describes the full functionality of a MuWNS-V. This new system consists of a Reference (tracking detector) using four MWPC chambers \cite{Varga_2016}, as well as a Receiver (comprising four, smaller and simpler MWPC-s). The two systems run two independent readouts based on Raspberry-Pi microcomputers, components which have been proven to be an efficient and reliable solution for other established muography projects such as imaging Sakurajima volcano \cite{Olah_2018} and for various underground structures \cite{Olah_2012} measurements. In the current work, in order to allow for flexibility and simplicity of this prototype, the Reference had an angular resolution of around 15 mrad, and a positional accuracy of 3.9 cm (RMS) was attained with the Reference located 2.4 meters above the Receiver. This can be improved in the future by employing the trackers with better angular resolution.

Both the Reference and the Receiver run independent internal clocks, which may be unstable on the level of a few ppm. Whenever a muon crosses both detectors, called "updates" \cite{Tanaka_vmups}, one can check if the detections happen within a certain time window. This is called the "verification time window", $T_W$. That is, any pair of detected particles, be it a valid muon or a background (two random particles or a shower) with time difference below $T_W$ will be defined as updates.

\section*{Results}

\subsection*{Experimental Setup}

\begin{figure}[ht]
\centering
\includegraphics[width=0.7\linewidth]{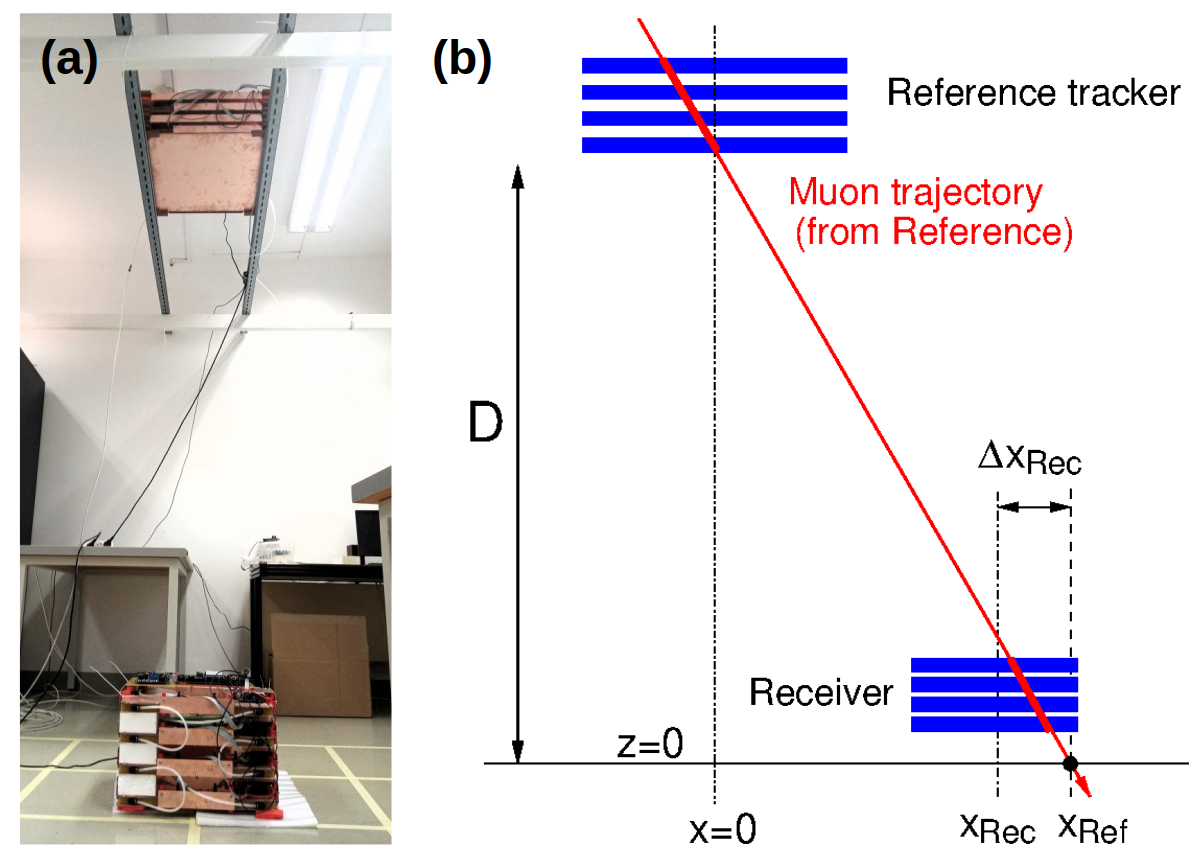}
\caption{The experimental setup. Photograph of the detector system (a), with Reference attached to rails below the ceiling, and the Receiver on the ground. An illustration (b) shows the geometrical parameters.}
\label{fig:vmups}
\end{figure}

The measurements were performed indoors, with the Reference tracker at $D = 2.4 m$ above the Receiver. Figure \ref{fig:vmups} shows a photograph of the structure, along with an illustration to clarify the geometrical parameters. The sensitive areas of the Reference, and the Receiver, are $S_{Ref} = 0.25 m^2$, and $S_{Rec} =0.145 m^2$ respectively (for muons passing vertically). The Receiver was moved in $x$ and $y$ directions on the floor. In the general configuration, the verification time window $T_W$ was set to $0.1 ms$.

The current experimental setup solves the problems which lies in the conventional muPS in the following way: 

(A) The positioning accuracy can be considerably improved without fancy atomic clocks by using directional information from the reference detector, in case that it includes a high precision and robust tracking system \cite{Gera_2022}.

(B) Cosmic muon events are not so frequent, and therefore, it is easy to find muons that cross both detectors without precise clocks or fast wireless communication ($T_W$ is practically in the order 0.1ms) \cite{Tanaka_vmups} to ensure, and continuously maintain time synchronization between the Reference and the Receiver. This concept called “Cosmic Timing Calibration” (CTC) \cite{Tanaka_ctc} and removes the necessity of wires and extraordinary fancy clocks from muPS. In this work, the CTC-based synchronization is maintained, or in other words, the two detectors are “CTC-locked” to each other.

\subsection*{Positioning error of a stationary Receiver}

In the experiment, both the Reference tracker and the Receiver were MWPC-based detector stacks, capable of determining the muon trajectory with a certain precision. Using either detectors, the muon trajectory was projected to the floor ($z=0$ plane), in order to determine the horizontal $(x,y)$ coordinates. The projected position from the fixed Reference is denoted by $(x_{Ref}, y_{Ref})$. One aims at estimating the unknown Receiver center position $(x_{Rec}, y_{Rec})$ with this configuration. The muon trajectory recorded in the Receiver pierces the $z=0$ plane at $(\Delta x_{Rec}, \Delta y_{Rec})$, measured relative to the Receiver center position (see Fig. \ref{fig:vmups}b).

When a muon passes through both the Reference and the Receiver (referred to as “updates” defined above), one can estimate the Receiver center position using the difference value between the projections:

\begin{equation}
(x_{Rec}, y_{Rec})= (x_{Ref}, y_{Ref}) - (\Delta x_{Rec}, \Delta y_{Rec})    
\end{equation}

The error of this position estimation is entirely dominated by the angular resolution of the Reference tracker, given the position resolution of both detectors is well below 1 cm. 

\begin{figure}[ht]
\centering
\includegraphics[width=0.7\linewidth]{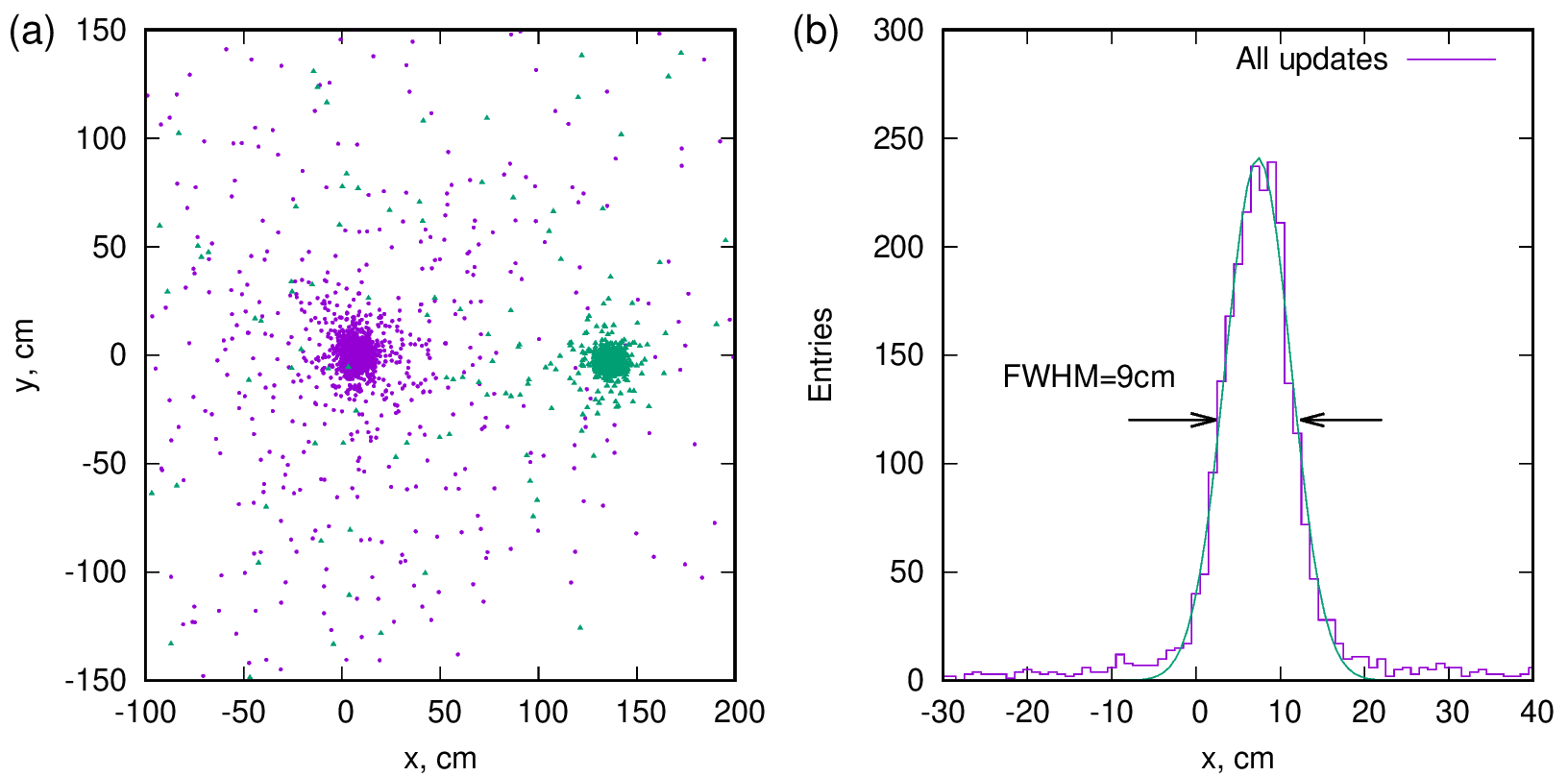}
\caption{Measured position resolution. Scatter plot of the estimated center points of the Receiver for all updates (a), at $T_W=0.1 ms$. Distribution of the estimated Receiver positions $x_{Rec} $ (b).}
\label{fig:positionerror}
\end{figure}

Figure \ref{fig:positionerror}a shows the scatter plot of the individual updates for two Receiver positions. With some background, the updates cluster around the fixed Receiver position. The position measurement error is 9 cm FWHM (3.9 cm RMS) shown in Fig. \ref{fig:positionerror}b.

\subsection*{Positioning a moving Receiver}

\begin{figure}[ht]
\centering
\includegraphics[width=0.7\linewidth]{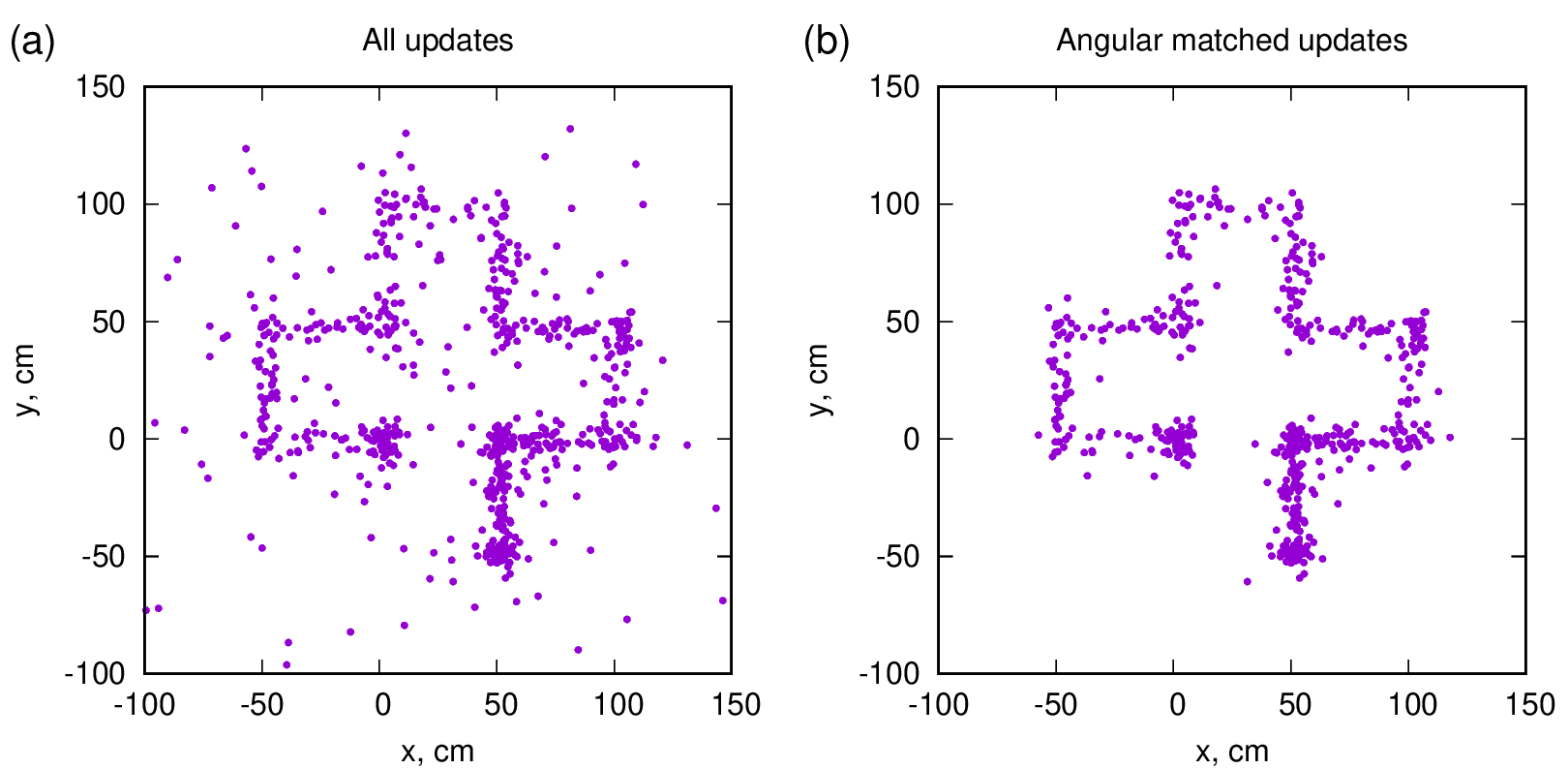}
\caption{Dynamic measurement done by moving the Receiver horizontally along grid lines on the floor. (a) Including all updates, some background is apparent, which is suppressed requiring a 80 mrad angular matching (b) between Receiver and Reference.}
\label{fig:angularmatch_moving}
\end{figure}

When the Receiver is moving, in this experiment on the $z=0$ plane of the floor, the updates continuously follow the position. For Figure \ref{fig:angularmatch_moving}, the Receiver was moved slowly along grid lines spaced $0.5 m$ apart. The pattern of the motion, starting from the origin, is clearly visible. The total path was $5 m$ long, taken in about 25 minutes. Figure \ref{fig:angularmatch_moving}a shows all the updates, with some background scattered around the real path. Figure \ref{fig:angularmatch_moving}b shows the same data, but requiring an angular matching between Receiver and Reference, as explained in the next section. 

\subsection*{Background rejection using the direction information in Receiver}

\begin{figure}[ht]
\centering
\includegraphics[width=0.7\linewidth]{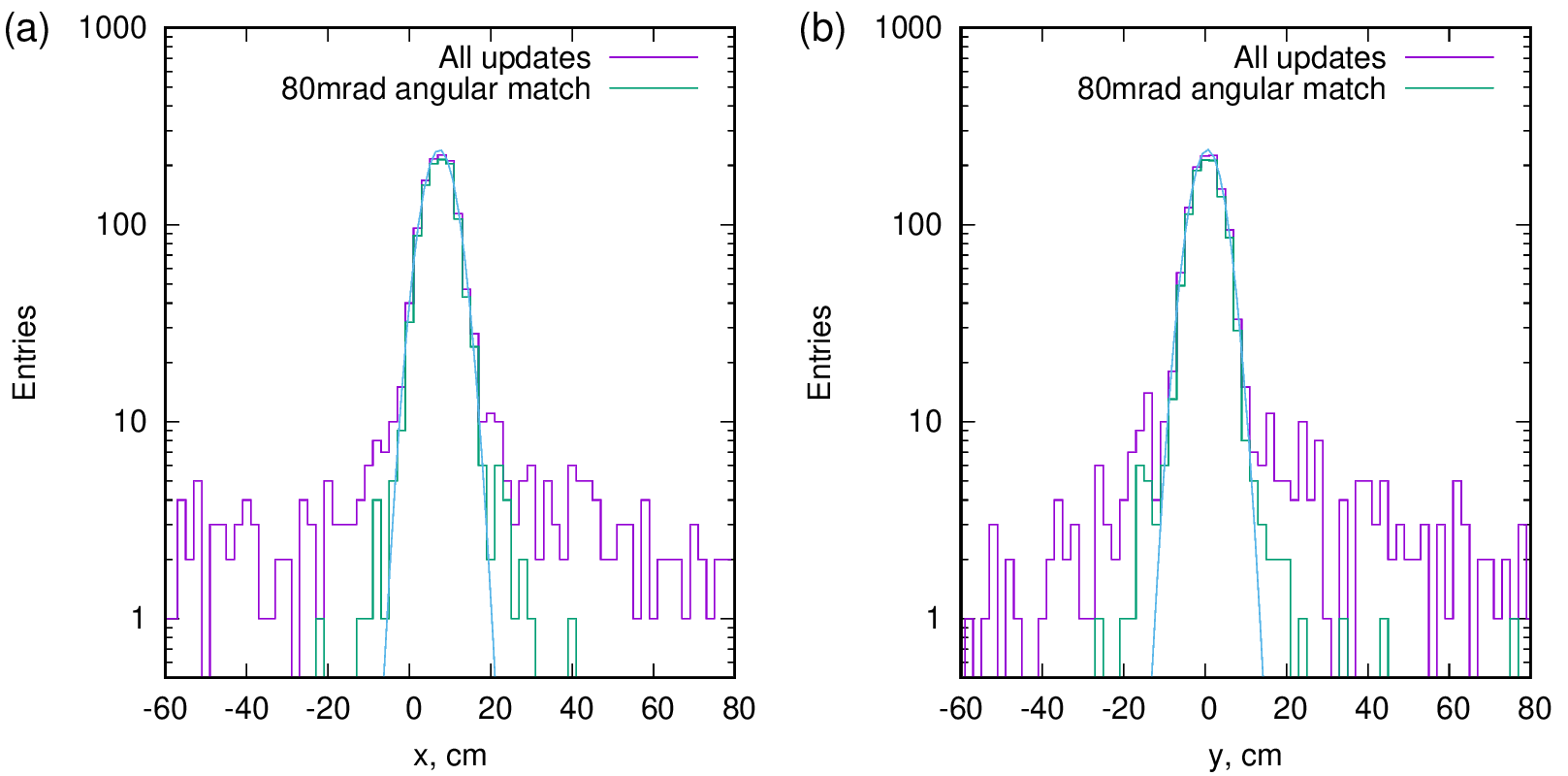}
\caption{Distribution of the positions from the updates, with or without using angular matching between Receiver and Reference tracker. The tails apparent for all updates get strongly suppressed when restricting the orientation in parallel directions, within 80mrad. Panels (a) and (b) show the x and y directions respectively.}
\label{fig:positionmatch}
\end{figure}

In a very typical configuration, the Receiver is a relatively small object, capable of only providing timing information. However, if the Receiver can measure a rough trajectory direction, the background may be reduced drastically \cite{Tanaka_vmups}. If one requests that the Reference and Receiver trajectories are parallel to each other within an 80 mrad (4.6 degrees) window, most background vanishes, as demonstrated in Fig.~\ref{fig:angularmatch_moving} right panel. The core of the distribution of the measured positions is unchanged, but the background populating the tails of the distribution is suppressed, as clearly apparent in Fig.~\ref{fig:positionmatch}. The source of this background is rather complicated and originates from air showers initiated by both primary cosmic rays as well as muon decays and other soft particles \cite{Olah_2017}.

\subsection*{Improvement of positioning accuracy by averaging}

\begin{figure}[ht]
\centering
\includegraphics[width=0.7\linewidth]{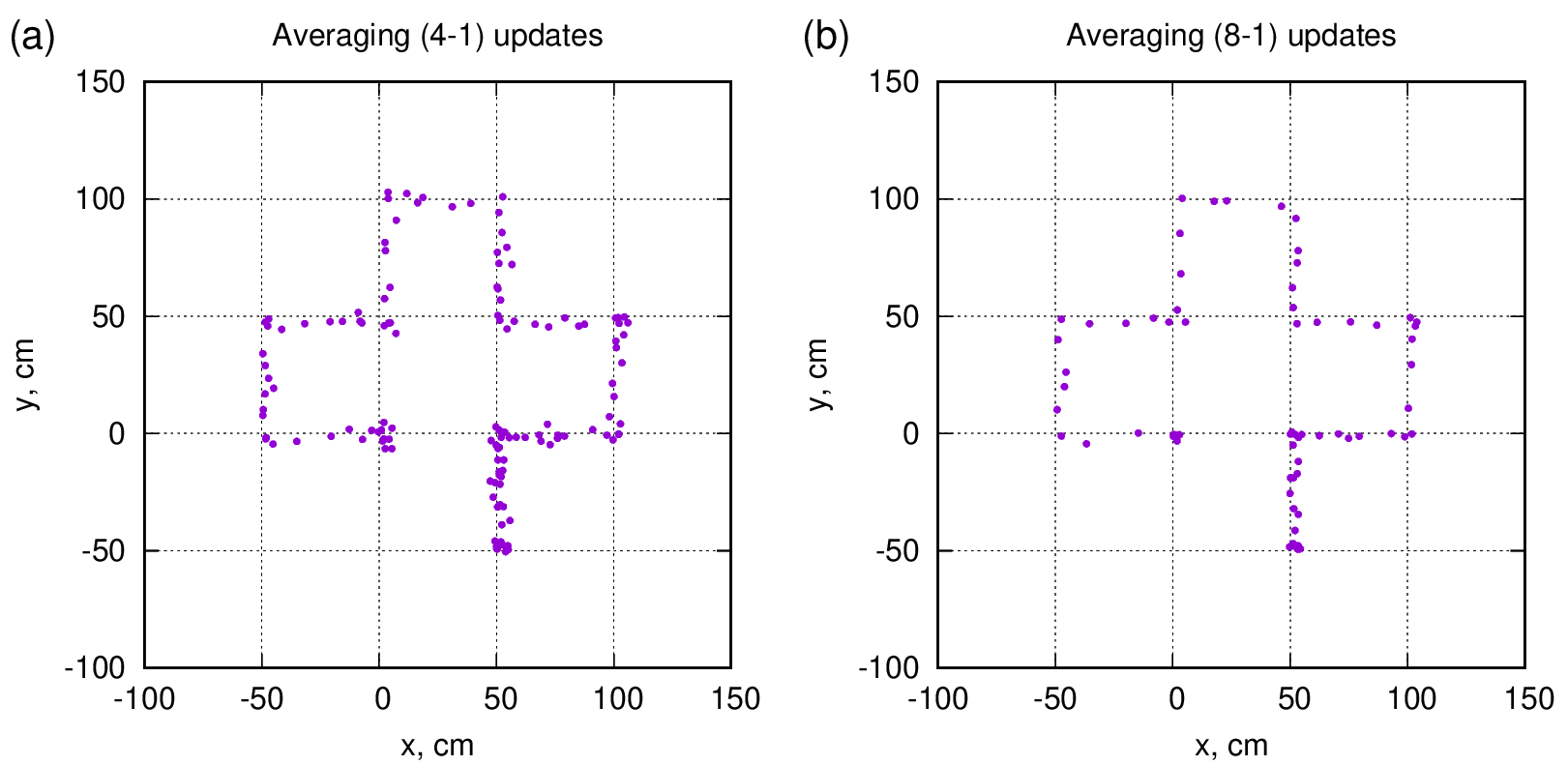}
\caption{Improvement of positioning accuracy by averaging over multiple updates. (a) averaging over “4-1” measurement points, that is, eliminating one out of 4 subsequent updates which are the furthest from the average. (b) the “8-1” average.}
\label{fig:average_moving}
\end{figure}

Since the Vector-muPS concept is statistical by nature, averaging over multiple update measurements gradually improves the positioning accuracy. This is demonstrated in Fig.~\ref{fig:average_moving}, which can be compared to Fig.~\ref{fig:angularmatch_moving}. A simple and consistent method to remove background outliers is to truncate the averaging: for Fig.~\ref{fig:average_moving}, one takes $N$ subsequent measurements, then takes the average, and then drops the one which is the farthest. Subsequently, the remaining $N-1$ are averaged. Figure~\ref{fig:average_moving}a shows $N=4$ and Fig.~\ref{fig:average_moving}b for $N=8$, where the positioning accuracy improvement is clearly visible, consistently with the $(N-1)^{-1/2}$ rule expectation.

\subsection*{Update frequency measurement}

The update frequency for vertical configuration was 0.60 $\pm$ 0.05 Hz, whereas it was measured to be 0.31 $\pm$ 0.05 Hz for 0.5rad (28 degrees) angle. The value for the vertical case can be readily estimated: assuming a vertical flux of $\phi_0 \approx 10^2 s^{-1}m^{-2}sr^{-1}$ indoors, the update rate $A$ is then obtained approximately as 

\begin{equation}
A= \phi_0 S_{Rec} (S_{Ref} D^{-2})    
\end{equation}

where $(S_{Ref} D^{-2})$ is the viewing solid angle of the Reference detector seen by the Receiver, with $S_{Rec}$ and $S_{Ref}$ being the areas of Receiver and Reference respectively. The calculation yields 0.64Hz, closely matching the measured value of 0.6Hz. At angles further from the zenith, the effective area of both Reference and Receiver decrease, in a complicated fashion due to the multiple layers of both tracking systems. There was no indication during the measurements that the update rate would deviate from the theoretical value, assuming the capture of nearly all muons with valid directional measurement.

\subsection*{Maintaining the CTC lock}

Given the fact that the Receiver can move anywhere and can also be anywhere, the only means to ensure that one can tag muons which cross both trackers (that is, a means of finding the updates) is sufficiently precise timing information. The readout systems of the two detectors are fully independent, which means that unless some synchronization is achieved, timing information is based on the internal clocks of the detectors. It is an inefficient solution to use exceedingly precise internal clocks.

Precise time synchronization can be naturally achieved using the CTC concept \cite{Tanaka_ctc}, as explained in the Introduction. Updates are not only for measuring position, but at the same time, are used to adjust the internal clocks – this can keep synchronization continuously, that is, the two systems are “locked” to each other. Once this CTC lock is captured, there is negligible chance to lose it (due to a sufficiently low level of random background).

For the measurement system presented above, the internal clocks of the two Raspberry-Pi microcomputers can be considered low precision. Even though this can be drastically improved using dedicated oscillators such as the oven controlled crystal oscillator (OCXO) \cite{Tanaka_wns2}, this still fits the demonstration purpose.

\begin{figure}[ht]
\centering
\includegraphics[width=0.7\linewidth]{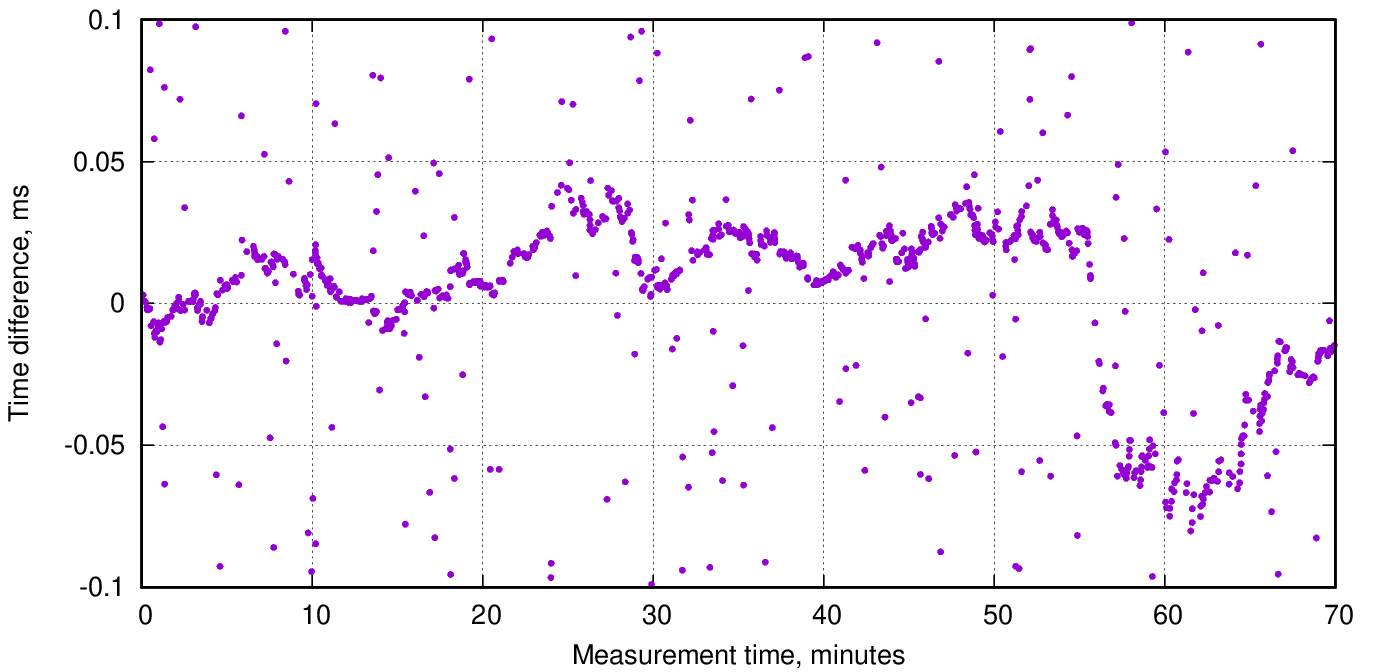}
\caption{CTC lock in practice. The vertical scale shows the time difference between a trigger time stamp in Receiver and Reference. The accumulation of points around zero corresponds to updates, whereas scattered points represent random background. A rolling average of 8 updates are used to maintain synchronization of the two fully independent internal clocks.}
\label{fig:ctc_lock}
\end{figure}

Figure \ref{fig:ctc_lock} shows how the “CTC lock” is maintained, for a time period of more than one hour. The update frequency was 0.31Hz for this data, that is, updates happen on average 3 seconds apart. After capturing 8 updates within the verification time window $T_W$  = 0.1 ms the Receiver clock is (off-line) adjusted to the Reference clock. The time difference between updates apparently drifts by 10-20 microseconds on the scale of minutes, which amounts to an order of a few ppm. Even with this low quality internal clock, and low update frequency, the CTC lock is safely maintained.

One must note that smart algorithms such as machine learning can considerably improve the verification of updates, by optimizing the conservative (in this case fixed) $T_W$ value, and exploiting constraints on the position of the Receiver (background points usually far away from the real position). In addition, for Fig. \ref{fig:ctc_lock}, angular matching was not used.

\subsection*{Capturing the CTC lock}

It has been demonstrated that once updates are verified, all subsequent updates can be captured to follow the motion and maintain synchronization. The question remains, how does one verify the first (few) updates unambiguously? In a possible practical configuration, such as an event of emergency, the Receiver is switched on such that its clock can be totally off from reality (by seconds), and its position is unknown. Below this essential function is demonstrated for the measurements in the present paper. Standard WiFi-like connection can easily adjust the clocks on the order of 0.1 seconds, whereas $T_W$  = 0.1 ms is fully sufficient to maintain the CTC lock; one must span this to 3 orders of magnitude in precision.  

\begin{figure}[ht]
\centering
\includegraphics[width=0.7\linewidth]{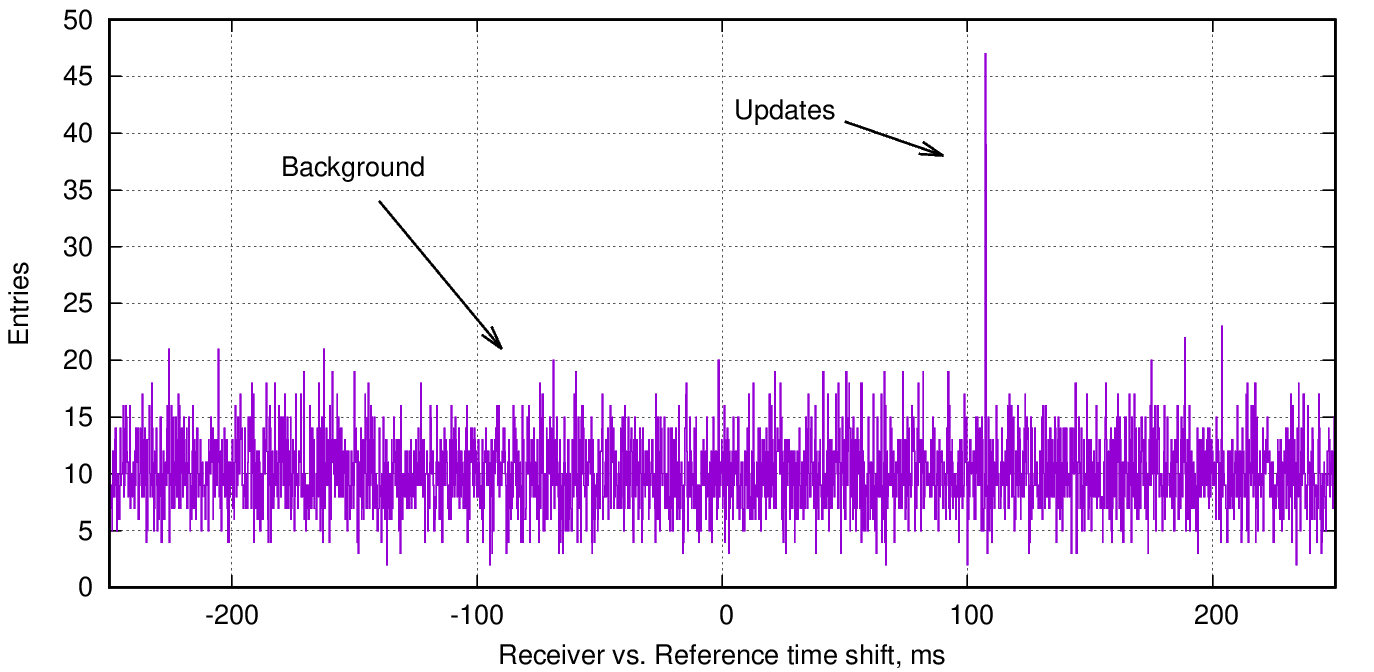}
\caption{As a function of the time shift $\Delta t$ between Receiver and Reference time stamps, the vertical axis shows the number of cases when the time difference is smaller than the verification time window $T_W$ between any combination of events (see text for clarification). The sharp peak corresponds to the real updates. All other values are from random background. Data taking time was 2 minutes.}
\label{fig:ctc_capture}
\end{figure}

In the first step, both Receiver and Reference take data for a few minutes (10-100 updates). Then the Receiver shares the time stamps with Reference, resulting in a time sequence $T_{Rec,1} \dots T_{Rec,N}$ . Accordingly, the time sequence for Reference is $T_{Ref,1} \dots T_{Ref,M}$  (both N and M are in the order of $10^4$ –- $10^5$). One can then evaluate for all combinations $(n,m)$ the time differences  $t_{diff}= (T_{Rec,n} - T_{Ref,m})$. The next step is to count cases when  $-T_W<(t_{diff}-\Delta t)< T_W$ . $\Delta t$ represents the unknown initial shift between the clocks, covering a typical range of 0.1 –- 1 second, for $\Delta t$ in steps of $2 T_W$. One expects a peak corresponding to the true $\Delta t$ value between Reference and Receiver: in that case, the true updates are adding up, whereas when $\Delta t$ is not the correct value, one gets only random (background) combinations. A typical such “time correlation” plot of the time sequences is shown in Fig.~\ref{fig:ctc_capture}, using data taken for only 2 minutes, at update frequency of 0.6Hz. In that case, the true time shift between the clocks was around 110ms. With this information, which is precise within an error of not more than $T_W$, the independent clocks can be initially adjusted. From then on, the CTC lock synchronization can be maintained for an arbitrarily long time.

\section*{Discussion}

The most recent iteration (current work) of MuWNS has 2 main elements; 1. Reference which is a good quality cosmic muon tracker with known position, and which points towards a smaller, less complex detector called Receiver. Every time a muon crosses both Reference and Receiver, the muPS PNT (positioning, navigation and timing) information is “updated” for the Receiver position and time synchronization between Reference and Receiver. A key point of the MuWNS-V is that Receiver and Reference can be strongly synchronized with a relatively slow wireless connection by using the CTC lock. With this CTC locking technique, even if the internal clocks of Receiver and Reference are not synchronized right after switching on these clocks, within the first several coincidence events, these two clocks are eventually synchronized by using muons that cross both Reference and Receiver. This approach is similar to the standard GPS systems, which require a few minutes (Time To First Fix) to give precise and continuous position information. In the current work, it was shown that CTC locking was achievable after capturing 10-100 coincidence events. This eliminates the need for highly sophisticated timing solutions. The results of the most recent experiment have demonstrated full operability of MuWNS-V under such conditions. Although accidental coincidence events have caused mis-positioning which could be resolved were by taking angular coincidence measurements between tracks at Reference and those at Receiver. MuWNS-V is most effective in situations when there is some connection (WiFi, Bluetooth, ultrasonic, etc.) available, but the position of the Receiver is unknown (e.g. GPS is unavailable). As described before, this may happen in underground or underwater conditions, or in a large building indoors, or in an emergency rescue situation with some overburden.

It is often advantageous to use two techniques in tandem to get the maximum benefit from both methods and to more effectively improve accuracy with calibration: such benefits are possible by combining muPS and Wi-Fi indoor positioning system (IPS). It is predicted that the global market value of Wi-Fi IPS will expand to 19 billion dollars by 2030 \cite{Mapsted_2023}. The clear benefit of Wi-Fi is, due to its widespread installation in buildings and its integration into smartphones, it can be nearly effortlessly retrofitted for the purpose of indoor positioning. The Wi-Fi positioning system (Wi-Fi IPS) offers indoor navigation capability for people with smartphones. Positioning accuracy of Wi-Fi IPS alone is limited to about 20 m without calibration; however, with calibration, Wi-Fi IPS can achieve 5-8 m accuracy. In order to attain further accuracy, various techniques such as Wi-Fi fingerprinting, ToF and AoA are combined to attain a sub-m positioning accuracy. Unlike other radio-wave-based IPSs, such as Bluetooth Low Energy (BLE), ZigBee, Radio Frequency Identification (RFID), Wi-Fi IPS does not require additional infrastructure since many of buildings already equipped with Wi-Fi access. Therefore, it is more reasonable to choose Wi-Fi IPS when the required positioning accuracy is in the order of one meter. This caveat is more or less the same for muPS as well: we need to deploy reference detectors in addition to Wi-Fi access points. Nevertheless, it is unusual that cm level navigation accuracy is required throughout an entire building. In many cases, we only need this level of accuracy in a localized area such as the IPS calibration points, the wireless power transfer (WPT) station for charging the autonomous robots, the area where articulated robot arms are located, etc. Working together in tandem, muPS offers positioning accuracy that is not affected by the surrounding environment and Wi-Fi IPS offers direct, practical communication via smartphones. For example, even if we sparsely deploy the muPS references on the ceiling of the indoor space, Wi-Fi IPS can be automatically calibrated when the clients are located near the muPS reference. 

\begin{figure}[ht]
\centering
\includegraphics[width=0.7\linewidth]{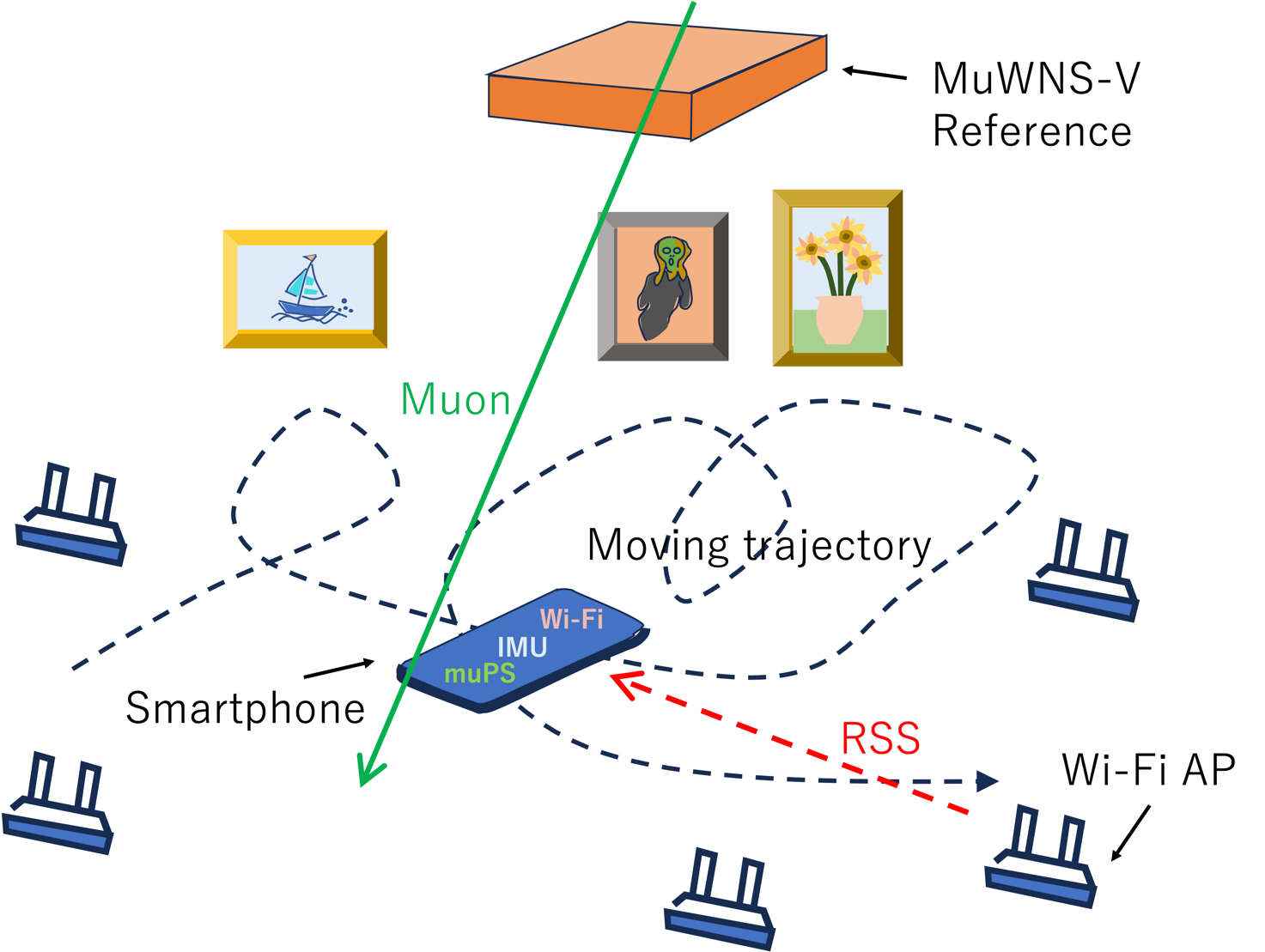}
\caption{Wi-Fi-muPS hybrid navigation scheme in a museum. Wi-Fi IPS is repeatedly calibrated with MuWNS-V.  RSS and IMU in this figure respectively stand for Received Signal Strength and Inertial Measurement Unit.}
\label{fig:museum}
\end{figure}

Unlike Wi-Fi IPS, clients cannot directly receive muPS signals with their smartphones in a practical way for direct utilization for real-time navigation. There are several reports claiming that muons could be detected by smartphone cameras by using the Complementary Metal-Oxide-Semiconductor’s (CMOS’s) silicon photodiode pixels to detect the photoelectric effect caused by muons \cite{Swaney_2021}. The detection efficiency is relatively high (~70\%) so, the muon events can be logged, stored, and compared with the events detected by the muPS reference. However, the size of these smartphone chips is usually too small to attain sufficient muPS signal update rate to perform real time navigation by itself. Since the typical size of these chips is $\approx 1 cm^2$, the vertical muons would only be detectable at approximately the timescale of once per minute. Although this muPS signal update rate is too low for navigation, muPS on regular smartphones could be used for occasional calibration of Wi-Fi IPS. Thus, indoor navigation accuracy and convenience could be improved with this approach. Wi-Fi-muPS hybrid navigation scheme is shown in Fig.~\ref{fig:museum}. In this use case, the clients are navigated inside a museum.  Each location and each artwork are tied in the navigation program, and clients are navigated to each artwork by this program. If the clients stay in the exhibition room for 10 minutes, their IPS in their smartphones will be calibrated by using MuWNS-V about 10 times before they leave the room. 

{\bf MuWNS-V-IMU hybrid navigation system.} In order to suppress the bias instability of the gyroscope and accelerometer as well as angle and velocity random walk errors, a ring laser gyroscope (RLG), a fiber optic gyroscope (FOG), or hemispherical resonator gyroscope (HRG) rather than a micro-electromechanical system (MEMS) gyroscopes are recommended. However, the weight of the device tends to be heavier when the former three options are employed. For example, the weight of MEMS is usually lighter than 500g, and the lightest model weighs only 1 g. However, the former three options weigh more than 500g, and this weight tends to increase as a function of the device stability improvement with some FOG models weighing almost 10 kg \cite{Borodacz_2022}. There is also a cost problem. The cheapest model of MEMS costs only 10 dollars but some FOG models cost 60,000 dollars; with the former option, gyroscope bias instability is more than 1,000 deg $hr^{-1}$ while the latter option suppresses it to less than 0.01 deg $hr^{-1}$ \cite{Borodacz_2022}. The situation with accelerometers is also more or less the same. The power consumption is also a large factor for practical navigation. While the power consumption of less accurate MEMS can be reduced to less than 100 mW, fancy FOG models require more than 20 W \cite{Borodacz_2022}. These factors make it very difficult to design a practical and stable stand-alone system for IMU navigation. MuWNS-V can be coupled with IMU in order to overcome the shortcomings of both methods. IMU short term drift can be repeatedly corrected by MuWNS-V, and a disadvantage of MuWNS-V (low positioning update rate) can be compensated by IMU.

{\bf MuWNS-V Security.} An IMU-based navigation device is self-contained thus, it can resist the influence of external disturbances well. In short, the location of an IMU is not identifiable by devices outside the system. Therefore, its navigation information cannot be jammed or spoofed. This is the major advantage against Wi-Fi IPS. While Wi-Fi is a widespread and well-known technology, it is highly susceptible to being cracked. Consequently, if utilized for navigation, malicious attackers could easily manipulate the Wi-Fi positioning information and as a result clients could be navigated to completely different locations. MuWNS-V used by itself can also be spoofed relatively easily since it uses Wi-Fi to communicate between the reference and the receiver. Potential attackers would only have to crack the system and provide incorrect reference tracking information to the receiver to interfere with its operation.  However, this problem could be solved by using MuWNS-V in conjunction with the Cosmic Coding and Transfer (COSMOCAT) technique \cite{Tanaka_coding}, \cite{Tanaka_coding2}. COSMOCAT could enhance the security of MuWNS-V with true-random-number-based cryptographic keys (generated from the arrival times of detected muons) which would encode the reference tracking information; each instance of tracking information could be encoded with a unique key that can be shared (without a physical connection) with the authentic receiver using the same keys automatically generated in the COSMOCAT process. By using COSMOCAT, the MuWNS-V-IMU hybrid navigation system could be immune against the threat of nearly all external cyber-attacks.

In conclusion, a new MuWNS-V prototype which is completely independent from preexisting navigation systems has been developed and demonstrated in this work. Unlike radio waves, acoustic signals, or laser beams, muometric positioning accuracy is not influenced by obstacles in its surrounding environment. The current muPS signal update rate was 0.3 – 0.6 Hz, but it can be improved by using larger reference trackers. Also, the current muometric positioning accuracy is 3.9 cm (RMS) which merely comes from the geometrical configuration (angular resolution) of the reference trackers. Therefore, it is anticipated that the current MuWNS-V prototype has a further potential to evolve towards sub cm real time positioning accuracy by attaining a muPS signal update rate of >1Hz.  This system was successfully developed and its performance has been confirmed to be sufficient for precise indoor navigation.

\section*{Methods}

The tracking systems for the purpose of this demonstration were MWPC detectors \cite{Varga_2016}, similar to those which proved long term reliable operation capability at the Sakurajima Muography Observatory \cite{Olah_2018}. For simplicity, only 4 layers were used, with 8 mm segmentation, resulting in a typical position resolution of about 5 mm. The Receiver was smaller with reduced 12 mm segmentation. 

The detector systems ran with fully independent Data Acquisition (DAQ) systems, which were based on a Raspberry Pi (model 3) microcomputer. The choice was motivated by its highly versatile nature, being similar to the capabilities of a standard smartphone, as well as its field-tested reliability. The two systems were taking data in parallel, but were started and stopped independently, and ensuring that there was no physical connection between the two – except for the cosmic muons.

The data contained individual muon trajectories, along with event time stamps, with the moderate precision and apparent drift as shown above. It seems to be straightforward strategy to improve the time resolution considerably, but it would not change the conclusions based on the measurements.

It is important to note that the data transfer rate was considerably low. Since the information from a single muon is merely around 50 bytes, with the Receiver running at 20Hz trigger rate, the necessary total data transfer rate is as low as 1kbyte/sec. 

\begin{figure}[ht]
\centering
\includegraphics[width=0.7\linewidth]{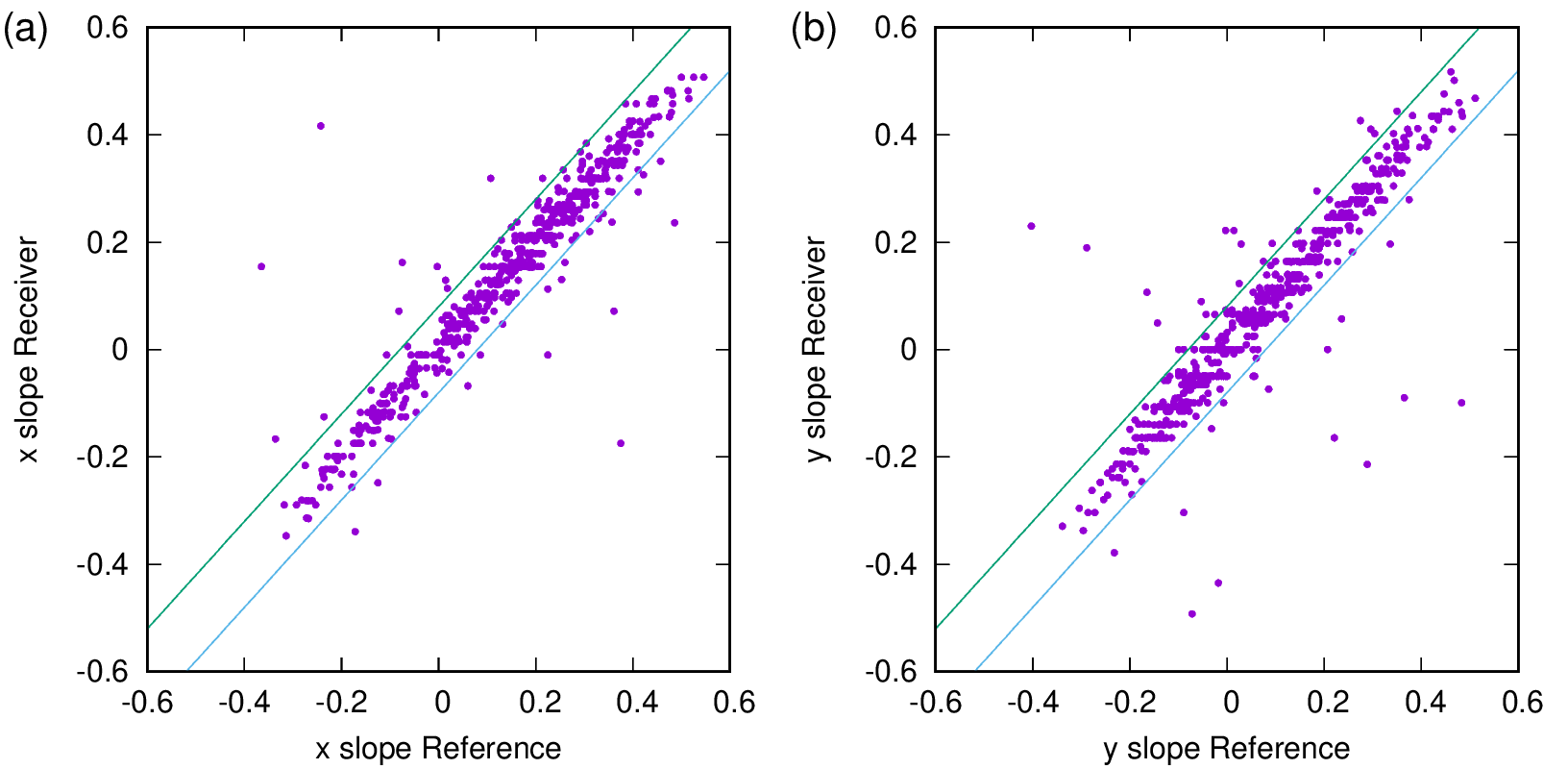}
\caption{Angular matching between Receiver and Reference. The figure shows the slopes (tangent of the angles) in x and y directions. The lines indicate the $\pm$ 80 mrad angular window which defines the angular matching.}
\label{fig:angularmatch_scatter}
\end{figure}

The Receiver provided information on the track direction, and based on the application, this may or may not be used. For angular matching, the Reference and Receiver tracks are closely parallel, as shown in Fig.~\ref{fig:angularmatch_scatter}. The angular resolution of the Receiver is much worse than that that of the Reference, however, 80 mrad captures most muons, and as it was demonstrated, helps in the task of reducing the random background.

% \section*{References}

\section*{Acknowledgements}

This work has been supported by the Hungarian NRDI Fund research grant under ID OTKA-FK-135349 and TKP2021-NKTA-10. Detector construction and testing was completed within the Vesztergombi Laboratory for High Energy Physics (VLAB) at Wigner RCP.

\section*{Author contributions statement}

Both authors conceived the experiments, wrote and reviewed the manuscript. D.V. conducted the experiments and prepared Figures 1 -- 7, and 9. HKMT prepared Figure 8.

\section*{Additional information}

\textbf{Competing interests} The authors declare no competing interests.

\textbf{Data availability} 

The datasets used and/or analyzed during the current study are available from the corresponding author on reasonable request.

\end{document}